\documentclass[11pt,epsf,letterpaper]{article}
\usepackage{subfig}
\usepackage{color}
\usepackage{amsmath}
\usepackage{amsfonts}
\usepackage{verbatim}
\usepackage{amssymb}
\usepackage{graphicx}
\usepackage{epstopdf}
\usepackage{mathrsfs}
\usepackage{epsfig}

\graphicspath{{./figs/}}
\providecommand{\U}[1]{\protect\rule{.1in}{.1in}}

\def\pa{\partial}

\renewcommand{\(}{\left(}
\renewcommand{\)}{\right)}
\renewcommand{\[}{\left[}
\renewcommand{\]}{\right]}

\begin{document}

	
\title{{\huge {Thermodynamics of asymptotically flat dyonic black holes}}}
\author{Ra\'ul Rojas Mej\'ias{}\footnote{Electronic address: \texttt{{\small raulox.2012@gmail.com}}} \\ \\ 
		\textit{Instituto de F\'\i sica, Pontificia Universidad Cat\'olica de Valpara\'\i so,} \\ 
		\textit{Casilla 4059,Valpara\'{\i}so, Chile.}\\[1mm]}
	\maketitle
	
	\begin{abstract} 
We investigate the thermodynamics of dyonic hairy black holes in flat spacetime when the asymptotic value of the scalar field is not fixed. We use the quasilocal formalism of Brown and York and corresponding boundary terms that make the variational principle well defined to prove that the scalar charges do not contribute to the first law of thermodynamics. We also provide a unified picture of obtaining exact solutions by comparing two different methods and discuss in detail the ansatz used in each coordinate system and the relation between them.
\end{abstract}

\newpage
\tableofcontents

\section{Introduction}
	
Einstein-Maxwell-dilaton theory appears naturally as a consistent truncation of low energy limit of string theory. The dilaton has a specific non-minimal coupling with the Maxwell term and the first exact hairy black hole solutions were constructed in \cite{Gibbons:1982ih},\cite{Gibbons:1987ps}, \cite{Garfinkle:1990qj}, and \cite{Ferrara:1995ih}.
Since in string theory the dilaton is a modulus whose expectation value controls the string dimensionless constant, $g_s$, it makes sense to understand how a variation of its asymptotic value affects the thermodynamics of hairy black holes. Such a study was performed in \cite{Gibbons:1996af} with a surprising result. That is, the first law of hairy black hole thermodynamics should be supplemented with contributions from the scalar fields.\footnote{A similar result was found for asymptotically AdS black holes; see \cite{Lu:2013ura,Lu:2014maa}.}

This result, which is based on implementing the Arnowitt-Deser-Misner (ADM) formalism \cite{Arnowitt:1959ah,Arnowitt:1960es} to hairy black hole thermodynamics, can be explicitly checked for various exact  solutions. However, from a physical point of view, the modification of the first law is puzzling. The scalar fields are characterized by charges that are not conserved, particularly in all known solutions there is no integration constant associated to the classical hair and that is why it is refereed to as `secondary hair'. Based on some earlier suggestions on how to solve this puzzle \cite{Astefanesei:2006sy,Astefanesei:2009wi}, Hajian and  Sheikh-Jabbari have used the phase space method to show in \cite{Hajian:2016iyp} that, indeed, the asymptotic value of the dilaton, $\phi_\infty$, is generally a redundant parameter. But if the non-conserved scalar charges are not allowed to appear in the first law, what exactly is inconsistent with the original proposal of \cite{Gibbons:1996af}? This question was completely answered in \cite{Astefanesei:2018vga}, where a general consistent variational principle was provided when the asymptotic value of the dilaton varies.\footnote{Another approach was also presented in \cite{Cardenas:2017chu} and, for asymptotically AdS spacetime, in \cite{Anabalon:2016ece} and \cite{Cardenas:2016uzx}.} In this case, the quasilocal energy computed by using the Brown-York formalism \cite{Brown:1992br} gets a new contribution due to the boundary term associated with the scalar field. This is consistent with the interpretation from string theory where a variation of the vacuum expectation value for the dilaton is equivalent with a change of the string coupling. This new method was explicitly checked for some exact electrically charged hairy black holes in \cite{Astefanesei:2018vga} and, later on, for several different examples in \cite{Naderi:2019jhn}. It is important to emphasize that the contribution from the scalar field to the conserved charges in asymptotically AdS spacetimes was clarified (see, e.g., \cite{Henneaux:2002wm,Barnich:2002pi,Henneaux:2004zi,Henneaux:2006hk,Anabalon:2015xvl}).

In this paper, we extend the work of \cite{Astefanesei:2018vga} to dyonic black holes. This is particularly important because, unlike the electrically charged case, for the dyonic black holes the extremal limit is well defined (see, e.g., \cite{Astefanesei:2019pfq} for a general discussion of different classes of asymptotically flat hairy black hole solutions) and so the canonical ensemble is well defined. We prove again, as expected, that the scalar charges do not appear explicitly in the first law as independent contributions. We also provide a detailed discussion on how two different methods can be used to obtain exact solutions and check the consistency of changing the coordinates from the frame used in \cite{Anabalon:2012ta,Anabalon:2013qua, Anabalon:2013sra,Anabalon:2015ija, Anabalon:2016izw} to the canonical frame \cite{Gibbons:1982ih} and \cite{Gibbons:1987ps}.

This paper is organized as follows. In Section \ref{sec:2}, we describe a method used to obtain exact solutions in a non-standard radial coordinate, and we employ it to reobtain well known solutions for both electrically and dyonic hairy black holes. We also provide, in Section \ref{subs:gs}, an example of one advantage of using this method in finding more general solutions. Also, we discuss the connection between the solutions obtained by this method and the ones obtained by using the canonical frame (with the usual radial coordinate). In Section \ref{sec:3}, we analyse the thermodynamic of these solutions when the asymptotic value of the dilaton is allowed to vary. We compute the on-shell action, the quasilocal stress tensor, the conserved energy, and verify the quantum statistical relation and first law of black hole thermodynamics. The analysis for the dyonic black hole in the Kaluza-Klein theory is left for the Appendix \ref{sec:kk}.

\section{Exact hairy black hole solutions}
\label{sec:2}

\subsection{Method and  setup}
\label{themethod}
In this section, we start by briefly describing the method proposed in \cite{Anabalon:2012ta,Anabalon:2013qua}, and also developed and used in 
\cite{Anabalon:2013sra,Anabalon:2015ija, Anabalon:2016izw,Acena:2012mr,Acena:2013jya,Anabalon:2014fla,Anabalon:2015xvl,Anabalon:2013eaa,Anabalon:2017yhv,Anabalon:2017eri}, to obtain exact black hole solutions. {Similar solutions were found using a different method in \cite{Gonzalez:2013aca,Martinez:2004nb}.
We are going to refer to this method and its corresponding coordinate system as the `$x-$frame'. 

We are interested in asymptotically flat hairy black hole solutions and the asymptotic value of the scalar field is not fixed in our analysis.
Let us consider the Einstein-Maxwell-dilaton action
\begin{equation}
I\[\,g_{\mu\nu},A_\mu,\phi\]=
\frac{1}{2\kappa}\int_{\mathcal{M}}
{d^4x\sqrt{-g}\[R-z(\phi)F^2 -\frac{1}{2}(\pa\phi)^2\]}
\label{action0}
\end{equation}
where $F^2\equiv F_{\mu\nu}F^{\mu\nu}$, $(\pa\phi)^2\equiv g^{\mu\nu}\pa_\mu\phi\pa_\nu\phi$, $R$ is the Ricci scalar, $F_{\mu\nu}=\pa_\mu A_\nu-\pa_\nu A_\mu$ represents the Maxwell (gauge) field and $z(\phi)$ is the coupling function between the scalar and gauge fields. In our conventions, $\kappa=8\pi$ and $G_N=c=1$.
The corresponding equations of motion are
\begin{align}
E_{\mu\nu}\equiv R_{\mu\nu}-\frac{1}{2}g_{\mu\nu} R
-\kappa\(T_{\mu\nu}^{EM}+T_{\mu\nu}^{\phi}\)
&=0,
\label{eins0}\\
\pa_\mu
\left[\sqrt{-g}z(\phi)F^{\mu\nu}\right]&=0,
\label{max0}\\
\frac{1}{\sqrt{-g}}\pa_\mu\(\sqrt{-g}g^{\mu\nu}\pa_\nu\phi\)
&=\frac{dz(\phi)}{d\phi}F^2
\label{klein0}
\end{align}
where the energy--momentum tensors for the gauge and scalar fields have the following expressions:
\begin{equation}
T_{\mu\nu}^{EM}=
\frac{2z(\phi)}{\kappa}
\[F_{\mu\alpha}F_{\nu}{}^{\alpha}-\frac{1}{4}g_{\mu\nu}F^2\], \quad T_{\mu\nu}^{\phi}
=\frac{1}{2\kappa}
\[\pa_\mu\phi\pa_\nu\phi-\frac{1}{2}g_{\mu\nu}
(\pa\phi)^2\]
\end{equation}

According to the $x-$frame method, the metric is initially put in the following static spherically symmetric ansatz
\begin{equation}
\label{ansatz1}
ds^2=\Omega(x)
\left[-f(x)dt^2+\frac{\eta^2dx^2}{f(x)}
+d\theta^2+\sin^2\theta d\varphi^2\right]
\end{equation}
where the coordinate $x$ plays the role of a radial coordinate,\footnote{The relation with the Schwarzschild radial coordinate is manifest in the asymptotic limit, where $\Omega(x)\approx r^2$.} and $\eta$ is introduced as a parameter of the solution. The following ansatz for the gauge field is provided
\begin{equation}
\label{Max2}
F\equiv \frac{1}{2}F_{\mu\nu}dx^\mu dx^\nu
=-\frac{q}{z(\phi)}dt\wedge dx+p\sin\theta d\theta\wedge d\varphi
\end{equation}
where $q$ and $p$ are the electric and magnetic parameters of the solution.
Note that both the ansatz (\ref{ansatz1}) and (\ref{Max2}) satisfy the Maxwell equations, (\ref{max0}). Now, by solving the combination of Einstein's equations $E^x_x-E^t_t$, one gets
\begin{equation}
\label{dilaton}
\phi'^2=
-\frac{2\Omega''}{\Omega}
+3\(\frac{\Omega'}{\Omega}\)^2
\end{equation}
which can be integrated provided an appropriate choice of $\Omega(x)$. Once $\phi=\phi(x)$ is known, the remaining equations of motion can be further integrated to get the metric function $f(x)$ and then the full solution. We will employ explicitly this method along the paper, and we will discuss the connection with the solutions obtained by using the canonical radial coordinate, which is given by a change of coordinates.

\subsection{General solutions}
\label{subs:gs}

As a concrete example of obtaining new regular solutions with the $x-$method presented in the previous subsection, let us consider a model with a non-trivial self-interaction for the scalar field --- the method of \cite{Gibbons:1982ih}, \cite{Gibbons:1987ps} and \cite{Garfinkle:1990qj} cannot be used in this case. Let us explicitly obtain
an exact solution for the case when the scalar field decays slowly enough up to a constant at the boundary, $\phi_\infty$. 
Let us first turn off the gauge field. Consider the following action:
\begin{equation}
I\[\,g_{\mu\nu},\phi\]=
\frac{1}{2\kappa}\int_{\mathcal{M}}
{d^4x\sqrt{-g}\[R-\frac{1}{2}(\pa\phi)^2-V(\phi)\]}
\label{actionV}
\end{equation}

The potential is `engineered' and, in this case, it has a quite simple expression
\begin{equation}
\label{potential}
V(\phi)=
\frac{2\alpha}{\nu^2}
\[\frac{\nu-1}{\nu+2}
\sinh{\frac{(\nu+1)\(\phi-\phi_\infty\)}{\sqrt{\nu^2-1}}}
-\frac{\nu+1}{\nu-2}
\sinh{\frac{(\nu-1)\(\phi-\phi_\infty\)}{\sqrt{\nu^2-1}}}
+\frac{4\nu^{2}-4}{\nu^{2}-4}
\sinh{\frac{\(\phi-\phi_\infty\)}{\sqrt{\nu^2-1}}}
\]
\end{equation}
where $\nu$ and $\alpha$ are two independent constants defining the theory. We emphasize that $\phi_{\infty}$ is a boundary condition and not an integration constant for a particular solution. {}{Since $\phi_\infty$ appears explicitly in the action (\ref{actionV}) through the potential for the scalar field, it must be regarded as a constant that is not allowed to vary, otherwise it might not be asymptotically Minkowski.}}

It was shown in \cite{Anabalon:2013qua} that, for $\phi_\infty=0$, a general family of solutions exists for this theory. Let us proceed by considering the ansatz for the metric given in equation (\ref{ansatz1}).
Now, the key point of this method is to choose a conformal factor such that the equation of  motion for the scalar field (\ref{dilaton}) can be integrated, which, according to \cite{Anabalon:2013qua}, is
\begin{equation}
\Omega(x)=
\frac {{\nu}^{2}{x}^{\nu-1}}{{\eta}^{2}
	\left({x}^{\nu}-1\right)^{2}}
\label{conf}
\end{equation}
We then obtain the following scalar field 
\begin{equation}
\phi(x)=\phi_\infty+\sqrt{\nu^2-1}\ln(x)
\label{scalar}
\end{equation}
which is consistent with the boundary condition. Finally, the metric function, after integrating the remaining equations of motion, is
\begin{equation}
f\left(x\right)
=\alpha\left[\left({\nu}^{2}-4\right)^{-1}-
{\frac {{x}^{2}}{{\nu}^{2}} \left(1+{\frac {{x}^{-\nu}}{\nu-2}}-
	{\frac {{x}^{\nu}}{\nu+2}}\right)}\right]
+{\frac {x{\eta}^{2}\left( {x}^{\nu}-1 \right) ^{2}}{{\nu}^{2}{x}^{\nu-1}}}
\end{equation}

The solution can be generalized by including a gauge field, so that the action becomes
\begin{equation}
\label{action1}
I\[\,g_{\mu\nu},A_\mu,\phi\]=
\frac{1}{2\kappa}\int_{\mathcal{M}}
{d^4x\sqrt{-g}\[R-e^{a\phi}F^2
	-\frac{1}{2}(\pa\phi)^2-V(\phi)\]}
\end{equation}
where $V(\phi)$ is the one given by (\ref{potential}), and $a$ is a coupling constant. The gauge potential is
\begin{equation}
A=\frac{qx^{-\nu}}{\nu^2}dt
\end{equation}
where $dF=A$ and $q$ is a charge parameter. Provided the ansatz (\ref{ansatz1}), the conformal factor (\ref{conf}) and the relation $a=(\frac{\nu-1}{\nu+1})^{1/2}$, one gets the same expression for the scalar field in (\ref{scalar}) and a more general metric function,
\begin{equation}
f\left(x\right)
={\frac{{\eta}^{2}{x}^{2}\left({x}^{\nu}-1\right)^{2}}
	{{\nu}^{2}{x}^{\nu}}}
+{\frac{\alpha}{{\nu}^{2}}
	\left({\frac{{x}^{\nu+2}}{\nu+2}}-{x}^{2}+{\frac{{x}^{2-\nu}}{2-\nu}}+{\frac {{\nu}^{2}}{{\nu}^{2}-4}}\right)}
-{\frac {{\eta}^{2}{x}^{2-2\,\nu}\left({x}^{\nu}-1 \right)^{3}}
	{2{\nu}^{3}\left(\nu-1 \right)}{\({q}^{2}
		{e}^{{\sqrt{\frac{{\nu}+1}{\nu-1}}\phi_\infty}}\)}}
\end{equation}

This solution admits two disconnected branches that can be analyzed independently. The branches correspond to separate spacetimes each one potentially containing a black hole, as we will explicitly show in the next subsections. {{}{In the remaining of the paper, we are going to consider only theories with $V(\phi)=0$, so that $\phi_\infty$ admits a variation.}}

\subsection{Electrically charged solution in the $x-$frame}
\label{sec:elec}

In the previous subsection, we showed, without a detailed derivation, that there are more general solutions which only can be obtained by using the $x-$frame. Here, we will use the method to reobtain, with more details, the very well known electrically charged black hole exact solutions with $V(\phi)=0$. These solutions were obtained first in \cite{Gibbons:1982ih}, \cite{Gibbons:1987ps} and \cite{Garfinkle:1990qj}. 

Let us turn off the magnetic charge, $p=0$, and consider the action (\ref{action0})
with $z(\phi)=e^{a\phi}$, where $a$ is an arbitrary coupling constant parametrizing the theory.\footnote{For the specific values $a=\{1, \sqrt{3}\}$, the model can be embedded in supergravity \cite{Gibbons:1982ih,Gibbons:1987ps,Garfinkle:1990qj,Ferrara:1995ih,Gibbons:1985ac,Dobiasch:1981vh}.} 
First, to conveniently decouple the equations of motion, let us separate the conformal factor as $\Omega(x)=\Omega_1(x)\tilde\Omega(x)$ and define a new coordinate $u\equiv\int{\tilde\Omega(x)dx}$. Then, by choosing  
\begin{equation}
\Omega_1(u)=\frac{1}{\eta^2(u-1)^2}
\end{equation}
the ansatz (\ref{ansatz1}) and (\ref{Max2}) can be put in the following form
\begin{align}
ds^2&=
\frac{1}{\eta^2(u-1)^2}
\[-{f(u)\tilde\Omega(u)}dt^2
+\frac{\eta^2du^2}{f(u)\tilde\Omega(u)}
+\tilde\Omega(u)
\(d\theta^2+\sin^2\theta \,d\varphi^2\)\] \label{a1}\\
F&=-\frac{qe^{-a\phi(u)}}{\tilde\Omega(u)}\, dt \wedge du \label{a2}
\end{align}

One aspect that can be observed at this level is that, in this coordinate system, the boundary of spacetime is located at $u=1$, where the conformal factor in the metric diverges (in the assumption that $\tilde\Omega$ remains finite at the boundary). The fact that $u$ can approach to 1 from the left or from the right side allows to recognize that there are two branches for this solution, corresponding to two disconnected spacetimes, one where $u$ ranges as $0<u<1$, which is called the negative branch, and other where $u>1$, called the positive branch.\footnote{In general, these branches have different properties and, then, the $x-$frame provides a new feature of the solutions. For instance, it was recently shown that when the scalar field is provided with a non-trivial self interaction in asymptotically flat theories, as shown in subsection \ref{subs:gs}, there are dinamically and thermodynamically stable black holes only in the positive branch \cite{Astefanesei:2019mds,Astefanesei:2019qsg}. Also, in the extended phase space of black hole chemistry \cite{Kubiznak:2014zwa,Kubiznak:2016qmn}, where the cosmological constant is considered to be a pressure term, there are non-trivial criticality phenomena in the positive branch for both canonical and grand canonical ensembles \cite{Astefanesei:2019ehu}.} 

Note, also, that the physical charge $Q$, up to a global sign, can be obtained by the Gauss Law, i.e., by integrating the Maxwell equation on the 2-sphere at infinity,\footnote{The convention is $\star F=\frac{1}{4}\sqrt{-g}\epsilon_{\mu\nu\alpha\beta}F^{\mu\nu}dx^\alpha\wedge dx^\beta$, where $\epsilon_{\mu\nu\alpha\beta}$ is the totally antisymmetric Levi-Civita symbol.}
\begin{equation}
\label{charg0}
Q
=\frac{1}{4\pi}\oint_{s^2_\infty}{e^{a\phi}\star F}
=\frac{1}{4\pi}
\oint{\sqrt{-g}e^{a\phi}F^{tu}d\theta\wedge d\varphi}=\frac{q}{\eta}
\end{equation}
Now, in order to solve the equations of motion, let us consider the combination $E_u^u-E_t^t$, which gives
\begin{equation}
\label{ttuu}
\phi'^2=\(\frac{\tilde\Omega'}{\tilde\Omega}\)^2
-\frac{2\tilde\Omega''}{\tilde\Omega},
\end{equation}
where prime symbol means $d/du$. The function $\tilde\Omega(u)$ can be chosen in two different ways, which defines the two families of solutions, similar to the ones in \cite{Zhou:2018ony}. We call \textit{family 1} to the family of solutions obtained by picking
\begin{equation}
\label{omeg1}
\tilde\Omega(u)
=\exp\[-a\(\phi-\phi_{\infty}\)\],
\end{equation}
and we call \textit{family 2} to the family obtained by picking
\begin{equation}
\label{omeg2}
\tilde\Omega(u)
=\exp\[\frac{1}{a}\(\phi-\phi_{\infty}\)\].
\end{equation}
where $\phi_\infty$ is the asymptotic value of the scalar field. Notice that $\tilde \Omega$ remains finite at the boundary, as commented earlier.

\textbf{Family 1:}
By integrating the equation (\ref{ttuu}), using (\ref{omeg1}), we obtain the following expression for the scalar field
\begin{equation}
\label{scalar1}
\phi(u)=\phi_\infty-\frac{2a}{1+a^2}\ln(u)
\end{equation}
and, by using (\ref{omeg1}) and (\ref{scalar1}), the remaining independent Einstein's equation can be integrated to get the last unknown metric function $f(u)$,\footnote{Note that the constants of integration were already assumed to be $\eta$ and $q$, therefore, any other constant appearing after integrating the equations of motion must be a suitable combination of $\eta$ and $q$.}
\begin{equation}
\label{metrf1}
f(u)=\(u-1\)^2u^{-\frac{3a^2-1}{a^2+1}}\eta^2
\[\(u-1\)\(1+a^2\)\(qe^{-\frac{1}{2}a\phi_\infty}\)^2+1\]
\end{equation}
Notice that $\lim_{u=1}{(-g_{tt})}=1$ and
$\lim_{u=1}{\phi(u)=\phi_{\infty}}$ as expected for asymptotic flatness and, as pointed out in \cite{Garfinkle:1990qj}, there is only one horizon,
\begin{equation}
\label{u1}
u_+=1-\frac{e^{a\phi_\infty}}{(1+a^2)q^2}
\end{equation} 
satisfying $h(u_+)=0$.

\textbf{Family 2:}
By integrating the equation (\ref{ttuu}), using (\ref{omeg2}), we obtain the following expression for the scalar field 
\begin{equation}
\label{scalar21}
\phi(u)=\phi_\infty+\frac{2a}{1+a^2}\ln(u)
\end{equation}
and, by using (\ref{omeg2}) and (\ref{scalar21}), the remaining independent Einstein's equation can be integrated to get
\begin{equation}
\label{metrf2}
f(u)=\(u-1\)^2u^{-\frac{4}{a^2+1}}\eta^2
\[-\(u-1\)\(1+a^2\)\(qe^{-\frac{1}{2}a\phi_\infty}\)^2+u\]
\end{equation}
which also satisfies $\lim_{u=1}(-g_{tt})=1$ and $\lim_{u=1}\phi=\phi_\infty$. There is, again, only one horizon
\begin{equation}
\label{solfam2}
u_+
=\frac{q^2e^{-a\phi_\infty}(1+a^2)}
{q^2e^{-a\phi_\infty}(1+a^2)-1}
\end{equation}
satisfying $h(u_+)=0$.

We would like to briefly comment on these two families. 
Let us follow the convention in which `negative branch' describes the domain \{$0<u<1$, $\phi<\phi_\infty$\}, and `positive branch' describes the domain \{$u>1$, $\phi>\phi_\infty$\}. By observing (\ref{scalar1}) and (\ref{scalar21}), it follows that, in order for the convention to be consistent, then family 1 should be associated with $a\leq 0$, and  family 2 is associated with $a>0$. Therefore, it is appropriate to say that, at least for the electrically charged black hole solutions without a potential for the scalar field, family 1 is adapted to the cases with $a\leq 0$, and family 2 to $a>0$. 

Finally, it is worth noticing that, for family 1, black hole configurations exist only in the negative branch, where the coupling function, $e^{a\phi}$, can take arbitrarily large values. On the other hand, for family 2, black hole configurations exist only in the positive branch, where, again, the coupling function can take arbitrarily large values.

\subsubsection{Comparison with the canonical frame}

Now, we would like to use a suitable change of coordinate in order to rewrite the solution in the canonical frame, as in \cite{Garfinkle:1990qj}, by using the standard radial coordinate, $r$. For concreteness, let us consider the family 1 (negative branch). 
The change of coordinates proposed is
\begin{equation}
\label{coord}
u=1-\frac{1}{\eta r}
\end{equation}
where $\eta$ is positive definite. By noting that $g_{uu}=g_{rr}\(dr/du\)^2$, the solution can be easily rewritten in the canonical radial coordinate as
\begin{equation}
ds^2=-a(r)^2dt^2+\frac{dr^2}{a(r)^2}
+b(r)^2\(d\theta^2+\sin^2\theta d\varphi^2\)
\end{equation}
with
\begin{equation}
a^2(r)\equiv 
\frac{(r-r_+)(r-r_0)^\frac{1-a^2}{1+a^2}}
{r^\frac{2}{1+a^2}} , \qquad
b^2(r)\equiv
r^2\(1-\frac{r_0}{r}\)^{\frac{2a^2}{a^2+1}}
\end{equation}
where $r_+=q^2e^{a\phi_\infty}(1+a^2)/\eta$ and $r_0=1/\eta$
are the black hole outer horizon and the location of the central singularity, respectively. To see that, at least for the cases $a<0$, $r_0$ is actually the central singularity, and not an inner horizon, observe that both the Ricci scalar and the scalar field diverge at the limit $r\rightarrow r_0$
\begin{equation}
R=\frac{2a^2(r-r_+)r_0^2r^{-\frac{2(a^2+2)}{a^2+1}}}
{\(a^2+1\)^2(r-r_0)^{\frac{1+3a^2}{1+a^2}}},
\qquad
\phi(r)=\phi_\infty
-\frac{2a}{1+a^2}\ln\(1-\frac{r_0}{r}\)
\end{equation}

It is convenient to introduce the so called `scalar charge' $\Sigma$ as the component in the subleading term in the asymptotic expansion of the scalar field in the canonical coordinate,  $\phi(r)=\phi_\infty+\frac{\Sigma}{r}+\mathcal{O}(r^{-2})$. In this case, the scalar field is expanded as
\begin{equation}
\phi(r)=\phi_\infty+\frac{2a}{(1+a^2)\eta r}
+\mathcal{O}(r^{-2})
\end{equation}
and, therefore, $\Sigma=\frac{2a}{(1+a^2)\eta}$.

Finally, notice that $a=0$ corresponds to the Reissner-Nordstr\"om limit. In that particular case, the scalar field becomes a constant, the Ricci scalar vanishes everywhere and $r_0=1/\eta$ becomes the inner horizon.

\subsection{Dyonic solution in the $x-$frame}
\label{sec:dyonic1}

In this subsection, we consider the case $z(\phi)=e^{\pm\phi}$ with both electric and magnetic charges turned on. We are going to reobtain the exact solution in \cite{Astefanesei:2006sy}, by using the method shown before. Let us, again, focus on the family 1 ($a=-1$),\footnote{The analysis is identical for the family 2 ($a=+1$).} and consider the ansatz
\begin{align}
ds^2&=\Omega(u)\[-f(u)dt^2+\frac{\eta^2 du^2}{u^2f(u)}
+d\theta^2+\sin^2\theta d\varphi^2\]  \\
F&=-\frac{qe^{\phi}}{u} \,dt \wedge du
+p\sin\theta\,d\theta\wedge d\varphi
\end{align}
where $\eta$, $q$ and $p$ are the three constants of integration (the parameters) of the solution. They are related with the physical electric and magnetic charges, $P$ and $Q$, which are
\begin{align}
\label{charges1}
Q=\frac{1}{4\pi}\oint_{s^2_\infty}{e^{-\phi}\star F}
=\frac{q}{\eta}, \qquad
P=\frac{1}{4\pi}\oint_{s^2_\infty}{F}=p\,,
\end{align}
respectively. The combination of the Einstein's equations $E_t^t-E_u^u$ leads to the equation
\begin{equation}
\label{comb1}
\phi'{}^2
=3\(\frac{\Omega'}{\Omega}\)^2
-\frac{2}{u}
\(\frac{\Omega''}{\Omega}+\frac{\Omega'}{\Omega}\)
\end{equation}
which, provided the conformal factor 
\begin{equation}
\Omega(u)=\frac{u}{\eta^2(u-1)^2}\,,
\end{equation}
can be integrated to obtain the following expression for the scalar field
\begin{equation}
\phi(u)=\phi_\infty+\ln(u)
\end{equation}
Now, the remaining independent equation of motion is solved by the following metric function
\begin{equation}
f(u)=\frac{\eta^2(u-1)^2}{u^2}
\[u+2u(u-1)\(qe^{\frac{1}{2}\phi_\infty}\)^2 
-2\eta^2(u-1)\(pe^{-\frac{1}{2}\phi_\infty}\)^2\]
\end{equation}

Unlike the electrically charged solution presented before, in this case, the horizon equation, $f=0$, implies the existence of two horizons,
\begin{equation}
u_{\pm}=
\frac{1}{2}+
\frac{2\eta^2e^{-\phi_\infty}p^2
	-1}{4q^2e^{\phi_\infty}}
\pm(\mp)\frac{
\sqrt{
4e^{-2\phi_\infty}p^4\eta^4
+4e^{2\phi_\infty}q^4
-2(2qp\eta)^2
-4\eta^2e^{-\phi_\infty}p^2
-4q^2e^{\phi_\infty}
+1}
}
{4q^2e^{\phi_\infty}}
\end{equation}
due to the magnetic charge.
The choice $\pm$ or $\mp$ depends on branch being considered. In negative branch ($0<u<1$), one must ensure that $u_{-}<u_{+}$ and, in positive branch ($1<u<\infty$), $u_+<u_-$.

\subsubsection{Comparison with the canonical frame}

Let us consider the same change of coordinate used for the electrically charged solution, (\ref{coord}). In the same manner, the solution can be rewritten as
\begin{equation}
ds^2=-a^2(r)dt^2+\frac{dr^2}{a^2(r)}
+b^2(r)\(d\theta^2+\sin^2\theta d\varphi^2\)
\end{equation}
where
\begin{equation}
a^2(r)=\frac{\(r-r_{+}\)\(r-r_{-}\)}{b^2(r)}, \qquad
b^2(r)=r\(r+\Sigma\)
\end{equation}
By asymptotically expanding the scalar field in the $r$ coordinate, one gets $\Sigma=-1/\eta$, and the inner and outer horizon, $r_{\pm}=1-\frac{1}{\eta u_\pm}$, can be written as
\begin{equation}
r_{\pm}=
\frac{\Sigma{q}^{2}}{e^{\phi_\infty}}
-\frac{p^2e^{\phi_\infty}}{\Sigma}
-\frac{\Sigma}{2}
\mp\frac{1}{2\Sigma}
\sqrt{
 \frac{4(q\Sigma)^4}{e^{2\phi_\infty}}
-8(qp\Sigma)^2
-{\frac{4(q\Sigma^2)^2}{{e^{\phi_\infty}}}}
+4(e^{\frac{1}{2}\phi_\infty}p)^4
-4(\Sigma e^{\frac{1}{2}\phi_\infty}p)^2
+\Sigma^4}
\end{equation}

Notice that $\Sigma$ is negative, according to our conventions. One can see that both the Ricci scalar and the scalar field diverge in the limit $r=-\Sigma$, which is the location of the central singularity,
\begin{equation}
R=
\frac{\Sigma^2\(r-r_{+}\)\(r-r_{-}\)}{2r^3(r+\Sigma)^3},\qquad
\phi(r)=\phi_\infty+\ln\(1+\frac{\Sigma}{r}\)
\end{equation}

Now, we are going to explore thermodynamic aspects of the solutions presented so far when the asymptotic value of the scalar field is not fixed.

\section{Thermodynamics and scalar charges}
\label{sec:3}

In this section, we are going to use the counterterm method and the quasilocal formalism for the asymptotically flat black hole solutions presented before, in order to compute the conserved energy and obtain the regularized on-shell action. This will allow to verify the quantum statistical relation and the first law of black hole thermodynamics, under the consideration that $\phi_\infty$ is allowed to vary.

It is well known that, once the model is embedded in string theory, $\phi_\infty$ is related to the string coupling; this makes clear that $\phi_\infty$ is a parameter that controls the theory rather than an integration constant specific to a particular solution (see, e.g., \cite{Hajian:2016iyp} and \cite{Astefanesei:2018vga} for some recent discussions in this context). In general relativity, the claim above may seem unusual because the boundary conditions are fixed. However, similar with the $AdS$ case \cite{Henneaux:2006hk,Anabalon:2015xvl,Hertog:2004ns,Caldarelli:2016nni}, first one expands all the fields at the boundary and, at this level, one impose boundary conditions that characterize a large set of solutions. Therefore, $\phi_\infty$ is a boundary condition, not an integration constant for a specific solution.

According with the variational principle when the asymptotic value of the scalar field is not fixed, the gravitational action for Einstein-Maxwell-dilaton theories (with $V(\phi)=0$), that consists of the bulk part of the action, the Gibbons-Hawking boundary term $I_{GH}$ and the gravitational counterterm $I_{ct}$, should be supplemented with a boundary term for the scalar field $I_\phi$, as
\begin{equation}
\label{totala}
I=I_{bulk}+I_{GH}+I_{ct}+I_\phi
\end{equation}
The gravitational counterterm that cancels infrared divergences in the theory, regularizing the action, is \cite{Lau:1999dp,Mann:1999pc,Kraus:1999di,Mann:2005yr} 
\begin{equation}
I_{ct}=-\frac{1}{\kappa}\int_{\pa\mathcal{M}}
{d^3y\sqrt{-h}{\sqrt{2\mathcal{R}^{(3)}}}},
\end{equation}
where $y^a=(t,\theta,\varphi)$ are the coordinates on the boundary $\pa\mathcal{M}$, which is the hypersurface $u=const$, and the general boundary term for scalar field, found in \cite{Astefanesei:2018vga}, is
\begin{equation}
I_{\phi}=-\frac{1}{2\kappa}\int_{\pa\mathcal{M}}
{d^3y\sqrt{-h}\[
\frac{(\phi-\phi_\infty)^2}{\Sigma^2}W(\phi_\infty)\]},
\label{scalarboundary}
\end{equation}
where the function $W$ is defined by means of the general boundary condition 
\begin{equation}
\Sigma\equiv\frac{dW(\phi_\infty)}{d\phi_\infty}
\end{equation}
which is similar with the one proposed in \cite{Hertog:2004ns} for AdS black holes.\footnote{A concrete relation $\Sigma=\Sigma(\phi_\infty)$ in not required and we can work in the general situation, $W=\int{\Sigma(\phi_\infty) d\phi_\infty}$. We remark that both $\phi_\infty$ and $\Sigma$ are coming from the asymptotic expansion of the scalar field which is independent of particular solutions.} The action (\ref{totala}) corresponds with the grand canonical ensemble, where $\delta A_t|_{\pa\mathcal{M}}=0$ is the boundary condition for the gauge potential. This corresponds to fixing the electric potential $\Phi\equiv A_t|_{\pa\mathcal{M}}-A_t|_{\text{horizon}}$.
The canonical ensemble, given by the boundary condition $\delta\left.\(z(\phi)\star F\)\right|_{\pa\mathcal{M}}=0$, which fixes the electric charge $Q$, is obtained by adding a new boundary term to the action,\footnote{The thermodynamic potentials of grand canonical ensemble and canonical ensemble are related by a Legendre transform in $Q$--$\Phi$.}
\begin{equation}
I_A=-\frac{2}{\kappa}\int_{\pa\mathcal{M}}{d^3y\sqrt{-h}z(\phi)n_\mu F^{\mu\nu}A_{\nu}}
\label{boundary}
\end{equation}
where $n_\mu$ is the normal unit to the boundary.

The quasilocal formalism of Brown and York \cite{Brown:1992br} provides a powerful method to obtain conserved quantities in general relativity. According to this method, the total energy of spacetime $E$ is the conserved quantity associated with the time-translational symmetry of the metric tensor, given by the Killing vector $\xi=\pa/\pa t$. If one consider a quasilocal surface with a stress tensor defined as
\begin{equation}
\tau_{ab}\equiv
\frac{2}{\sqrt{-h}}\frac{\delta I}{\delta h^{ab}}
\end{equation}
where $I$ is the total action given by (\ref{totala}), then, the conserved energy is
\begin{equation}
\label{energy}
E=\oint_{s^2_\infty}
{d^2\sigma\sqrt{\sigma}n^a\tau_{ab}\xi^b}
\end{equation}
where $d^2\sigma=d\theta d\varphi$ (for the spherical cross section) and $n_a=(-g^{tt})^{-1/2}\delta{_a ^t}$ is the normal unit to the hypersurface $t=const$ at the asymptotic limit. The concrete expression for the quasilocal stress tensor for the total action including the gravitational counterterm was found in \cite{Astefanesei:2005ad} and it
is given, in this case, by\footnote{Concrete applications of this method for asymptotically flat black holes can be found in \cite{Astefanesei:2009wi,Astefanesei:2006zd,Astefanesei:2009mc,Astefanesei:2010bm}}
\begin{equation}
\tau_{ab}
=\frac{1}{\kappa}\[K_{ab}-h_{ab}K-\Psi
\(\mathcal{R}^{(3)}_{ab}-\mathcal{R}^{(3)}h_{ab}\)
-h_{ab}\Box\Psi+\Psi_{;ab}\]
+\frac{h_{ab}}{2\kappa}
\frac{(\phi-\phi_\infty)^2W}{\Sigma^2}
\end{equation}
where $\Psi\equiv\left(\frac{1}{2}\mathcal{R}^{(3)}\right)^{-1/2}$. The fact that the function $W(\phi_\infty)$, coming from general boundary conditions for the scalar field, appears in the expression for the quasilocal stress tensor indicates that it may contribute to the total energy, as we shall verify in the next cases. 

\subsection{Electrically charged solutions}

In this subsection, we are going to consider the electrically charged solutions from subsection \ref{sec:elec}. Concretely, the family 1. We will compute the conserved energy, obtain the regularized Euclidean on-shell action and verify the quantum statistical relation.

From the trace of Einstein's equation, (\ref{eins0}), it follows that $R=\frac{1}{2}\(\pa\phi\)^2$ and, therefore, the bulk part of the action is
\begin{equation}
I_{bulk}^E
=-\frac{q^2\beta}{2\eta}
\int_{u_+}^{u_b}
{\frac{du}{\tilde\Omega(u)e^{a\phi}}}
=\frac{\beta q^2e^{-a\phi_\infty}}{2\eta}(u_+-1)
\end{equation}
where $u_b$ is the coordinate at the boundary, which finally should be pushed to $u_b\rightarrow 1$. $\beta$ is the periodicity in the imaginary time ($\tau^E=-it$) that removes the conical singularity in the Euclidean version of the metric, and it is the inverse of the Hawking temperature, $\beta=T^{-1}$.
The Gibbons-Hawking boundary term has the following fall-off
\begin{equation}
I_{GH}^E
=\frac{\beta}{4\eta}\[3(a^2+1)q^2e^{-a\phi_\infty}
+\frac{a^2+3}{a^2-1}\]
+\frac{\beta}{\eta(u_b-1)}
+\mathcal{O}(u_b-1)
\end{equation}
By noting that the Ricci scalar of the foliation $u=const$ is $\mathcal{R}^{(3)}=2\eta^2(u-1)^2/\tilde\Omega$, we obtain the fall-off of the gravitational counterterm,
\begin{equation}
I_{ct}^E
=-\frac{\beta}{2\eta}
\[1+(a^2+1)q^2e^{-a\phi_\infty}\]
-\frac{\beta}{\eta(u_b-1)}
+\mathcal{O}(u_b-1)
\end{equation}
and, finally, the boundary term of the scalar field has a finite contribution at the boundary
\begin{equation}
I_{\phi}^E=-\frac{\beta}{4\eta}
{\tilde\Omega(u_b)\[f(u_b)\tilde\Omega(u_b)\]^{1/2}}
\frac{\(\ln{u_b}\)^2}{(u_b-1)^3}W
=\frac{1}{4}\beta W
\end{equation}
Therefore, the total on-shell action at the limit $u_b=1$ is
\begin{equation}
I^E=I_{bulk}^E+I_{GH}^E+I_{ct}^E+I_{\phi}^E
=\beta\[
\frac{(a^2+2u_{+}-1)}{4\eta}q^2e^{-a\phi_\infty}
-\frac{1}{4\eta}\(\frac{a^2-1}{a^2+1}\)
+\frac{1}{4} W\]
\label{onshell1}
\end{equation}

In order to idenfity $I^E$ with the thermodynamic potential, let us now obtain the expressions for the thermodynamic quantities of this solutions.
The Hawking temperature $T$, the Bekenstein-Hawking entropy $S$, the electric charge $Q$ and the conjugate potential $\Phi$ are
\begin{align}
T&=\frac{\tilde\Omega(u_+)}{4\pi\eta}
\left.\frac{df(u)}{du}\right|_{u_+}
=\frac{\eta(u_{+}-1)^2u_+^{-\frac{a^2-1}{a^2+1}}}{4\pi u_+}
\[\frac{2(a^2+2u_{+}-1)}{(a^2+1)(u_{+}-1)}
+(3a^2+4u_{+}-1)q^2e^{-a\phi_\infty}
-1\]\\
S&=\frac{\pi\tilde\Omega(u_+)}{\eta^2(u_{+}-1)^2},
\qquad
Q=\frac{q}{\eta},\qquad
\Phi=-qe^{-a\phi_\infty}(u_{+}-1),
\end{align}
while the conserved energy is obtained by computing the quasilocal stress tensor, which, in this case, has the following expansion
\begin{equation}
\tau_{tt}
=\frac{1}{\kappa}
\left\{
\[\frac{a^2-1}{a^2+1}-(a^2+1)q^2e^{-a\phi_\infty}\]\eta-\frac{1}{2}\eta^2W\right\}(u_b-1)^2
+\mathcal{O}\[\(u_b-1\)^3\]
\end{equation}
With this result, the conserved energy is
\begin{equation}
E=\frac{1}{2\eta}\[(a^2+1)q^2e^{-a\phi_\infty}
-\frac{a^2-1}{a^2+1}\]+\frac{1}{4}W
\end{equation}
which mathches with the ADM mass $M$, obtained by expanding $g_{tt}$, in the limit $W=0$, that is, when $\phi_\infty$ is considered fixed from the beginning.
We conclude that the conserved energy is $E=M+\frac{1}{4}W$, where\footnote{Notice that the factor $1/4$, appearing in this expression, does not appear in \cite{Astefanesei:2018vga}. This is due to the convention used in the action for the scalar field kinetic term.}
\begin{equation}
M=\frac{1}{2\eta}\[(a^2+1)q^2e^{-a\phi_\infty}
-\frac{a^2-1}{a^2+1}\]
\end{equation}
It is worth noting that the quantity added to the ADM mass is only proportional to the asymptotic value of the scalar field in the particular case where $\Sigma$ does not depend on $\phi_\infty$. In general, $W$ is not proportional to $\phi_\infty$.

It is easy now to verify that the both the quantum statistical relation and the first law of black hole thermodynamics
\begin{equation}
I^E=\beta\(E-TS-Q\Phi\)\equiv \beta\mathcal{G} ,
\qquad dE=TdS+\Phi dQ
\end{equation}
where $\mathcal{G}$ is the thermodynamic potential for grand canonical ensemble, hold.
In canonical ensemble, where the electric charge is fixed, the boundary term (\ref{boundary}) contributes as $I_{A}^E=\beta Q\Phi$ and the quantum statistical relation reads
$\tilde I^E=\beta(E-TS)$, where $\tilde I^E=I^E+I_A^E$.

Finally, notice that, despite the electrically charged solution has not an extremal limit well defined, for the dyonic solution, one can use the entropy function formalism \cite{Sen:2005wa,Astefanesei:2006dd} 
to show that the extremal limit is allowed \cite{Astefanesei:2006sy}.

\subsection{The dyonic solution, $a=-1$}

Now, we would like to perform a similar analysis in the presence of a magnetic charge, for the case $a=-1$. We proceed in the same manner that the previous subsection.
Note, however, that the gauge potential can be written as
\begin{equation}
A=\(\int{\frac{qe^\phi}{u}du}+C_1\)dt+\(-p\cos\theta+C_2\)d\varphi
\end{equation}
where $C_1$ and $C_2$ are additive constants. This implies that, if we consider the boundary condition $\left.\delta A_t\right|_{\pa\mathcal{M}}=0$, the contribution at the boundary coming from the variation of the action is
\begin{equation}
\delta I =-\frac{2}{\kappa}\left.\int{d^3x\sqrt{-g}e^{\alpha\phi}F^{\theta\varphi}\delta A_\varphi}\right|^{\theta=\pi}_{\theta=0}
=\frac{2}{\kappa}\left.\int{d^3x
\(\frac{p\eta e^{-\phi}}{u}\)
\delta A_\varphi}\right|^{\theta=\pi}_{\theta=0}
\end{equation}
Let us choose the additive constant $C_2$ in the expression for the gauge potential in the following way: inside the domain $0<\theta<\frac{\pi}{2}$ we take $C_2=-P$ and inside $\frac{\pi}{2}<\theta<\pi$ we take $C_2=P$. In this way, we can integrate from $\theta=0$ to $\pi/2-\epsilon$ and from $\pi/2+\epsilon$ to $\pi$, avoiding the Dirac string as long as we take the limit $\epsilon\rightarrow 0$. In the Euclidean section, we have then
\begin{equation}
\delta I
=-\frac{2}{\kappa}\left.\int{d^3x
	\(\frac{p\eta e^{-\phi}}{u}\)\delta A_\varphi}\right|^{\theta=\frac{\pi}{2}-\epsilon}_{\theta=0}
-\frac{2}{\kappa}\left.\int{d^3x
	\(\frac{p\eta e^{-\phi}}{u}\)\delta A_\varphi}\right|^{\pi}_{\theta=\frac{\pi}{2}+\epsilon}
=\beta\Psi^p\delta{p}
\end{equation}
where $\Psi^p\equiv -\int_{u_+}^{u=1}{\frac{p\eta e^{-\phi}}{u}du}$ is the conjugate (magnetic) potential. In order to have a well-defined action principle, one need to add to the action an extra term coming from the magnetic sector. In the grand canonical ensemble, this extra term is going to appear in the Euclidean action as $I_p^E=-\beta P\Psi^p$.

The total on-shell action is, therefore,
\begin{equation}
I^E=I_{bulk}^E+I_{GH}^E+I_{ct}^E+I_{\phi}^E+I_p^E
=\beta\[\frac{q^2e^{\phi_\infty}u_+}{2\eta}
-\frac{\eta p^2 e^{\phi_\infty}\(2u_{+}-1\)}{2u_+}
-P\Psi^p+\frac{1}{4}W\]
\end{equation}
Now, by computing the relevant component of the quasilocal stress tensor, whose asymptotic expansion gives
\begin{equation}
\tau_{tt}=\frac{1}{\kappa}
\[2\eta\(- q^2e^{\phi_\infty}
+\eta^2p^2e^{-\phi_\infty}\)
-\frac{1}{2}\eta^2 W\]
(u_b-1)^2+\mathcal{O}\[\(u_b-1\)^3\]
\end{equation}
we can obtain the total energy, in the limit $u_b=1$, which is
\begin{equation}
E=\frac{q^2e^{\phi_\infty}}{\eta}
-\eta p^2e^{-\phi_\infty}+\frac{1}{4}W
=M+\frac{1}{4}W
\end{equation}
where $M$ is the ADM mass.

Now, let us verify the quantum statistical relation.
The Hawking temperature and entropy are
\begin{align}
T&=\frac{u_+}{4\pi\eta}
\left.\frac{dh(u)}{du}\right|_{u_+}
=\frac{\eta\(u_+-1\)^2}{2\pi u_+^2}
\[
u_+\(2u_{+}+1\)q^2e^{\phi_\infty}
-\(u_{+}+2\)\eta^2p^2e^{-\phi_\infty}
-\frac{u_+\(u_{+}+1\)}{2(u_{+}-1)}\] \\
S&=\pi\Omega(u_+)
\end{align}
The electric and magnetic charges are given in (\ref{charges1}) and the conjugate potentials, the electric $\Phi$ and magnetic one $\Psi^p$, are
\begin{equation}
\Phi=-\(u_{+}-1\)qe^{\phi_{\infty}},\quad
\Psi^p\equiv A^p_t(u=1)-A^p(u_+)
=-\frac{\eta\(u_{+}-1\)}{u_+}pe^{\phi_\infty}
\end{equation}
where, in the language of differential forms, $dA^p=F^p\equiv e^{-\phi}\star F$ or, equivalently, $\Psi^p=\int_{u=1}^{u_+}{F^p_{ut}du}$.
One can see that indeed both the quantum statistical relation and the first law,
\begin{equation}
I^E=\beta\(E-TS-Q\Phi-P\Psi^p\), \qquad
dE=TdS+\Phi dQ+\Psi^p dP
\end{equation}
hold, as expected, without including explicitly the `scalar charge' $\Sigma$. These results support the conclusions of \cite{Astefanesei:2018vga}.

\section{Conclusion}

In this work, we use the method developed in \cite{Anabalon:2013qua} to re-obtain some known exact hairy black hole solutions of Einstein-Maxwell-dilaton asymptotically flat theories \cite{Garfinkle:1990qj}. However, we have generalized the method to obtain solutions when the asymptotic value of the scalar field $\phi_\infty$ is not fixed, which provides an interesting tool for constructing this type of exact solutions. We have obtained two different families of solutions and each of them with two different branches or domains of the radial coordinate. We have also analyzed some specific aspects of black hole thermodynamics, using the quasilocal formalism, supplemented with the boundary terms that make the action principle well defined when the asymptotic value of the scalar field varies.

{{}{It is important to remark that we analyzed two different kind of theories, one with $V(\phi)\neq 0$ where $\phi_\infty$ is a constant that cannot be varied and other with $V(\phi)=0$, which allows variations of $\phi_\infty$. We realize that, regarding $\phi_\infty$, the conserved energy matches the ADM mass for any value of $\phi_\infty$ in the former and for $\phi_\infty=0$ in the latter.}}

With respect to thermodynamics, we first analyzed hairy electrically charged black holes for more slightly general theories than considered previously in \cite{Astefanesei:2018vga} for asymptotically flat spacetimes, concretely, for arbitrary value of the constant in the coupling function $e^{a\phi}$ between the scalar and Maxwell fields. 
We checked that the conserved energy receives a new contribution coming from the asymptotic value of the scalar field, $E=M+\frac{1}{4}W$, where $W=W(\phi_\infty)$ is defined by the relation between $\Sigma$, appearing in the subleading term in the asymptotic expansion of the scalar field, and $\phi_\infty$.
This might be compared with AdS, when one considers a scalar field with the fall-off
$\phi(r)={\alpha}/{r}+{\beta}/{r^2}+\mathcal{O}(r^{-2})$
where the conserved energy reduces to $M$ when either $\alpha=0$, $\beta=0$, or under a non-trivial third condition $\beta\propto \alpha^2$, that is not present in the flat case. We observe that once the correct total energy is computed, there is no need of including the scalar charges in the first law of thermodynamics.

One observation, not new in the context of \cite{Garfinkle:1990qj}, is that hairy (with no self-interacting potential) electrically charged asymptotically flat exact black hole solutions in four dimensions do not have a well defined extremal limit. In this limit, the solution becomes a naked singularity. 
This is particularly problematic in the scenario when we do thermodynamics with $Q$ fixed, since, as it was pointed out in \cite{Chamblin:1999hg}, a vacuum solution with a fixed $Q\neq 0$ is not a regular solution of Einstein's equations and, therefore, cannot be chosen as the ground state of the theory. A natural candidate that can be used as a background for the canonical ensemble is, then, the extremal black hole and so the importance of constructing solutions with such a limit becomes evident.
By turning on the magnetic charge, we obtain regular extremal hairy dyonic black hole solutions.
Consequently, the canonical ensemble can be appropriately defined. In this context, we verified that the first law is indeed satisfied without including the extra term
due to the non-conserved scalar charges.
Therefore, we have generalized the results presented in \cite{Astefanesei:2018vga} to the dyonic sector.

{{}{As a final comment, we would like to point out that the present analysis is not the most general one. In order to understand these solutions and their thermodynamics on a deeper level, a general analysis in the lines of \cite{Henneaux:2018hdj,Henneaux:2020ekh} is required when scalar fields are turned on.}}

\section*{Acknowledgments}

I would like to thank my supervisor Dumitru Astefanesei for suggesting this problem and valuable guidance for completing this project. I also would like to thank the anonymous referees for their important remarks and suggestions.
The work was partially supported by the national Ph.D. scholarship Conicyt 21140024.

\newpage

\appendix
\section{The dyonic solution, $a=-\sqrt{3}$}
\label{sec:kk}

Since this solution is algebraically more complicated in this case, let us consider the well known exact solution \cite{Gibbons:1985ac} as written in \cite{Astefanesei:2006sy}. The metric is
\begin{align}
ds^2=-a(r)^2dt^2+\frac{dr^2}{a(r)^2}
+b(r)^2\(d\theta^2+\sin^2\theta d\varphi^2\)
\end{align}
\begin{equation}
a(r)^2
=\frac{\(r-r_{+}\)\(r-r_{-}\)}
{\sqrt{A(r)B(r)}}, \qquad
b(r)^2=\sqrt{A(r)B(r)}, \qquad
\phi(r)=\phi_\infty
+\frac{\sqrt{3}}{2}
\ln\[\frac{A(r)}{B(r)}\]
\end{equation}
where
\begin{equation}
\label{horpm}
r_{\pm}\;=\;M\pm c, \qquad 
c=\frac{1}{2}\sqrt{4M^2+\Sigma^2
	-4q^2e^{\sqrt{3}\phi_\infty}-4p^2e^{-\sqrt{3}\phi_\infty}}
\end{equation}
and
\begin{equation}
A(r)=\(r-r_{A_+}\)\(r-r_{A_-}\),\qquad
B(r)=\(r-r_{B_+}\)\(r-r_{B_-}\)
\end{equation}
with
\begin{align}
\label{AABB}
r_{A_\pm}=-\frac{\Sigma}{2\sqrt3}\;\pm
pe^{-\frac{1}{2}\sqrt{3}\phi_\infty}\sqrt{\frac{2\Sigma}{\Sigma+2\sqrt{3}M}},\qquad
r_{B_\pm}=\;\frac{\Sigma}{2\sqrt3}\;\pm
qe^{\frac{1}{2}\sqrt{3}\phi_\infty}\sqrt{\frac{2\Sigma}{\Sigma-2\sqrt{3}M}}
\end{align}
and the constraint
\begin{equation}
\label{const}
\frac{1}{6}\Sigma=\frac{p^2e^{-\sqrt{3}\phi_\infty}}
{\Sigma+2\sqrt{3}M}+
\frac{q^2e^{\sqrt{3}\phi_\infty}}
{\Sigma-2\sqrt{3}M}
\end{equation}
The gauge field, on the other hand, is
\begin{equation}
F=\frac{qe^{\sqrt{3}\phi}}{b^2}dt\wedge dr
+p\sin\theta d\theta\wedge d\varphi
\end{equation}

Now, let us schematically show the connection between the solution in canonical frame, as given above, with the $x-$frame. The change of coordinates that connects them is
\begin{equation}
\label{finalchange}
u=\(\frac{A}{B}\)^{\pm 1}
\end{equation}
so that the scalar field can be rewritten in the $u$--coordinate as 
\begin{equation}
\phi(u)=\phi_\infty\pm\frac{\sqrt{3}}{2}\ln\(u\)
\end{equation}
The choice of $\pm$ sign defines the two families of the solution in $x-$frame, according to the previous discussions. If, by considering the change of coordinates (\ref{finalchange}), we define $\Omega(u)=u^{\pm\frac{1}{2}}B(u)$, then, the metric can be rewritten as
\begin{equation}
ds^2=\Omega(u)
\[-h(u)dt^2+\frac{\eta^2du^2}{f(u)}+d\theta^2+\sin^2\theta d\varphi^2\]
\end{equation} 
where $h(u)$ and $f(u)$ must be determined in order to complete the relation between the $x-$frame and the canonical frame.

\subsection{Thermodynamics}

To obtain the Euclidean action, we proceed as before. Note that, from Einstein equations we have $R=\frac{1}{2}\(\pa\phi\)^2$ and, from Klein-Gordon equation, (\ref{klein0}), we have $\sqrt{3}e^{-\sqrt{3}\phi}F^2=-
\frac{1}{\sqrt{-g}}\(a^2b^2\sin^2\theta\phi'\)'$. Therefore, the bulk part of the action is
\begin{equation}
I_{bulk}^E=
-\frac{\beta}{4\sqrt{3}}
\left.\(a^2b^2\phi'\)\right|^{r_b}_{r_+}
=\frac{1}{4\sqrt{3}}\beta\Sigma+\mathcal{O}(r_b^{-1})
\end{equation}
The Gibbons-Hawking boundary term, the gravitational counterterm and the boundary term for the scalar field are
\begin{equation}
I_{GH}^E=\frac{3}{2}\beta M-\beta r_b+\mathcal{O}(r_b^{-1})
,\qquad
I_{ct}^E=\beta r_b-\beta M+\mathcal{O}(r_b^{-1})
,\qquad
I_{\phi}^E=\frac{1}{4}\beta W+\mathcal{O}(r_b^{-1})
\end{equation}
and the total action, considering the contribution from the magnetic sector, is then
\begin{equation}
I^E=I_{bulk}^E+I_{GH}^E+I_{ct}^E+I_{\phi}^E+I_p^E
=\beta
\(\frac{\Sigma}{4\sqrt{3}}+\frac{1}{2}M-P\Psi^p+\frac{1}{4}W\)
\end{equation}
By computing the conserved energy by using the quasilocal formalism, we get
\begin{equation}
E=-a^3b
\[ab'-1-\frac{1}{4\Sigma^2}b\(\phi-\phi_\infty\)^2W\]
=M+\frac{1}{4}W
\end{equation}
as expected. 
Finally, to verify the quantum statistical relation, let us obtain the thermodynamic quantities. The Hawking temperature and entropy, in the compact form, can be written as
\begin{equation}
T=\frac{r_{+}-r_{-}}{4\pi\sqrt{A(r_+)B(r_+)}},\qquad
S=\pi\sqrt{A(r_+)B(r_+)}
\end{equation}
The electric and magnetic charges are $Q=q$ and $P=p$, and the concrete expressions for electric and magnetic potentials can be written as
\begin{equation}
\Phi=
\frac{36MQe^{\sqrt{3}\phi_\infty}}
{\(\sqrt{3}\Sigma-6{r_+}\)\(\sqrt {3}\Sigma-6M\)}
,\qquad 
\Psi^p=
\frac{36MPe^{-\sqrt {3}\phi_\infty}}
{\(\sqrt{3}\Sigma+6M\)\(\sqrt{3}\Sigma+6{r_+}\)}
\end{equation}
respectively. One can now easily verify that, by using the expression (\ref{horpm}), (\ref{AABB}) and (\ref{const}) for this solution, both the quantum statistical relation and the first law
\begin{equation}
I^E=\beta(E-TS-Q\Phi-\Psi^pP), \qquad dE=TdS+\Phi dQ+\Psi^p dP
\end{equation}
hold, once again, with no scalar charge $\Sigma$ appearing explicitly.

\end{document}